\documentclass{optica-article}

\journal{opticajournal} 

\articletype{Research Article}

\usepackage{lineno}

\begin{document}

\title{Zeroth-order-free holographic reconstruction with a nanoimprinted nonlocal metasurface}

\author{Teruyoshi Nobukawa,\authormark{1,*} Shunsuke Murai,\authormark{2,$^\dagger$}, Ryo Higashida\authormark{1} Yuta Yamaguchi\authormark{1}, Masato Miura\authormark{1}, Koichi Okamoto\authormark{2}, and Nobuhiko Funabashi\authormark{1}}

\address{\authormark{1}Science \& Technology Research Laboratories, Japan Broadcasting Corporation NHK, 1-10-11 Kinuta, Setagaya-ku, Tokyo 157-8510, Japan\\
\authormark{2}Department of Physics and Electronics, Graduate School of Engineering, Osaka Metropolitan University, 1-1 Gakuencho, Naka-ku, Sakai, Osaka, 599-8531, Japan}

\email{\authormark{*}nobukawa.t-eq@nhk.or.jp} 
\email{\authormark{$\dagger$}murai@omu.ac.jp}


\begin{abstract*} 
The undesired zeroth-order diffraction (ZOD) arising from imperfections in diffractive optical elements (DOEs) degrades the quality of target optical wavefronts.
Herein, we propose a zeroth-order-free holographic reconstruction method using a nanoimprinted nonlocal metasurface.
By judiciously designing the metasurface structure and its angular selectivity based on guided mode resonance, the ZOD can be suppressed without relying on a bulky, conventional 4\textit{f} setup.
We designed and fabricated a nanoimprinted nonlocal metasurface using a high-refractive-index TiO\textsubscript{2}-composite resin.
Using the metasurface, we demonstrated ZOD suppression in a surface‑relief DOE and a spatial light modulator.
Furthermore, we prototyped a 20‑mm‑square metasurface and verified its effectiveness in suppressing the ZOD in 3D holographic projection.
\end{abstract*}

\section{Introduction}
Diffractive optical elements (DOEs) modulate the phase of light to generate arbitrary optical wavefronts via propagation and diffraction.
Owing to this capability, they have been widely utilized across various optics and photonics fields, including microscopy \cite{Maurer2011}, optical tweezers \cite{Gieseler2021}, laser processing \cite{Malinauskas2016,Kumagai2017}, structured light \cite{Forbes2021}, holographic displays \cite{Blinder2022,Jiang2019,Yamaguchi2024,Higashida2024}, and computational imaging \cite{Heide2016,Wang2025,Nobukawa2025,Gao2026}.
Typically, DOEs are implemented using substrates featuring surface relief structures or engineered refractive index distributions that introduce wavelength scale optical path differences, or by employing a spatial light modulator (SLM) \cite{Khonina2024,Lazarev2019}.
Alternatively, wavefront generation can also be achieved using metasurfaces composed of subwavelength nanostructures \cite{Yu2011}, which allow not only phase manipulation but also control over amplitude, polarization \cite{Mueller2017}, and even dispersion \cite{Chen2020}.
A common issue shared by the above-mentioned optical elements and devices when generating wavefronts is the presence of undesired zeroth order diffraction (ZOD), which degrades the quality of the target wavefront.
The undesired ZOD arises from imperfections in DOEs.
In surface‑relief elements and metasurfaces, it is mainly caused by fabrication errors in the structure depths or in the geometry of the meta‑atoms.
In the case of SLMs, ZOD originates from pixel‑to‑pixel gaps, surface reflections, non-uniformity, and fringing‑field effects \cite{Palima2007,Moser2019,Engstrom2013}.
Furthermore, mismatches between the refractive index assumed in the optical design and that of the actual material, as well as deviations between the design wavelength and the wavelength used in the optical setup, also give rise to ZOD.

The common approach to suppress ZOD is a use of 4\textit{f} setup, as described in the following paragraph.
In parallel, in the fields of computer‑generated holograms, DOEs, and SLMs, several approaches have been investigated to suppress the ZOD.
One approach is to superimpose an off‑axis carrier phase or a defocus phase onto the phase pattern used to generate the target wavefront.
By introducing such additional phase terms, the ZOD can be spatially separated from the target wavefront either laterally in the transverse plane or axially along the optical axis near the Fourier plane \cite{Zhang2009}.
Another approach compresses the phase‑level range of a phase pattern used to produce the target wavefront to a narrower interval than 0–2$\pi$ which produces an additional zeroth-order beam \cite{Liang2012}.
By adjusting the amplitude of this beam to match that of the ZOD and adding a bias phase term to ensure that its phase becomes opposite to that of the ZOD, destructive interference can be induced.
In this way, the ZOD can be suppressed while the desired target wavefront is maintained.
A further method involves accurately measuring the characterization of an SLM and calibrating it, which eliminates the ZOD \cite{Engstrom2013}.
In addition, feedback‑based optimization techniques have been proposed, in which the diffracted field is captured by an image sensor and the phase distribution is iteratively updated to minimize the ZOD \cite{Matsumoto2012,Choi2021}.
However, these techniques inevitably sacrifice usable spatial‑frequency bandwidth or optical efficiency to avoid ZOD, and require redesigning or recalculating the phase distribution for each target wavefront.
While the above techniques rely on numerical redesign or algorithmic optimization of the phase distribution, approaches that incorporate auxiliary optical components have also been investigated for ZOD suppression.
One such method employs a checkerboard phase mask, which shifts the ZOD component into higher‑spatial‑frequency regions through the mask’s alternating phase structure \cite{Wong2008}.
Another reported technique utilizes volume filters, where the wavelength or angular selectivity of a 1D multilayer dielectric mirror is exploited to suppress the ZOD \cite{Bang2019}.
However, the former approach requires highly accurate alignment of the checkerboard phase mask and also necessitates recalculating the DOE phase pattern.
The latter approach, on the other hand, involves multiple film‑deposition steps, and its angular selectivity becomes ring‑shaped due to the one‑dimensional multilayer structure, resulting in limited flexibility in designing the filter characteristics.

The simplest and most widely adopted traditional approach is the use of a 4\textit{f} setup.
Because the ZOD is generally close to a plane wave, it is focused onto the optical axis by the Fourier transform of the first lens.
The focused spot corresponding to the ZOD is then blocked by a spatial filter or a mask, and the second lens performs a Fourier transform, yielding a high‑quality wavefront without ZOD.
However, the 4\textit{f} setup inevitably increases the size of the optical setup, posing a major obstacle for compact implementations.
Furthermore, the two lenses in the 4\textit{f} setup introduce bandwidth limitations and lens aberrations.
These issues become even more severe when generating large‑angle diffraction patterns from fine‑pitch DOEs, metasurfaces, and SLMs.
Although such aberrations can be mitigated by employing sophisticated compound lenses, doing so further increases the overall system size and complexity.

Meanwhile, in the field of metasurfaces, numerous studies have proposed thin and flat optics that replace conventional lenses and filters.
Among them, nonlocal metasurface refers to the metasurface where the optical response comes not solely from the local elements but from extended and collective modes\cite{RN3075,RN3076,RN3062,RN3063,Shastri2022,Pearson2025}. This nonlocality gives additional degree of freedom in designing the metasurface, and also robustness to the fabrication imperfection.
They have been used to implement functionalities such as edge detection \cite{Chamoli2025}, free space compression \cite{Guo2020}, wavelength filtering \cite{Song2021}, Zernike phase‑contrast imaging \cite{Ji2022}, and nonlinear up-conversion \cite{ValenciaMolina2024}.

In this study, we propose a zeroth-order-free holographic reconstruction method using a nonlocal metasurface.
The nonlocal metasurface used in this study operates on the basis of guided mode resonance (GMR) \cite{Huang2023,Quaranta2018}.
By exploiting the angle selectivity inherent to GMR structures, the ZOD can be effectively suppressed.
The proposed method requires only placing a single lightweight metasurface, with a thickness of a few hundred nanometers, behind the DOE/SLM, thereby eliminating the need for a bulky, conventional 4\textit{f} optical system.
In addition, the proposed method requires neither recalculation of the DOE or SLM phase distribution nor high‑precision alignment.
Consequently, ZOD suppression can be achieved with a much simpler optical configuration, free from lens‑induced aberrations.
In addition, no recalculation of the phase distribution or system calibration is required.
For manufacturability and scalability, we fabricated the nonlocal metasurface using a TiO\textsubscript{2}-composite resin by nanoimprint lithography.
While previous studies have demonstrated the fabrication of high‑aspect‑ratio metasurfaces using nanoimprint lithography \cite{Oh2026,Yoon2020,Miyata2022}, the present work focuses on fabricating a simple line‑and‑space structure for GMR.
A common challenge in nanoimprint processes is the presence of a residual layer after imprinting \cite{Dirdal2020, Park2025}.
However, GMR metasurfaces require a uniform layer beneath the nanostructures to support the guided mode, and we show that the residual layer naturally produced by nanoimprinting can serve this role without additional processing.

The remainder of this paper is organized as follows.
Section 2 introduces the principle and concept of the proposed method.
Section 3 describes the design, fabrication, and characterization of the nonlocal metasurface.
Section 4 demonstrates the application of the metasurface to a checkerboard diffraction grating and to SLM‑based holographic reconstruction for experimental validation.
In addition, we prototyped a 20 $\times$ 20 mm$^2$ metasurface and applied it to a 3D holographically generated wavefront, demonstrating its capability in a practical holographic projection scenario.
Finally, Section 5 concludes the paper.

\section{Principle and concept}

\begin{figure}[b]
\centering\includegraphics[width=12cm]{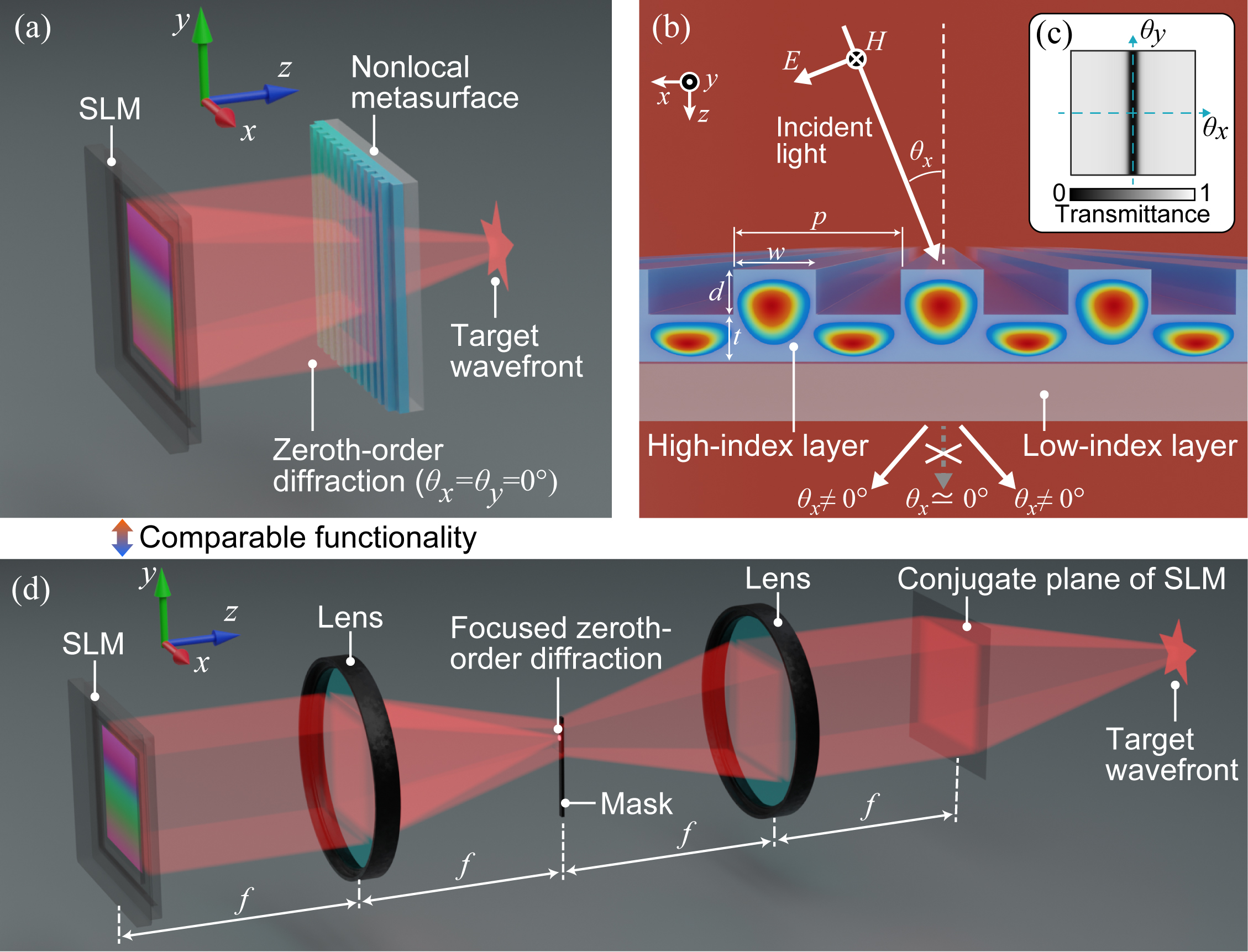}
\caption{Schematic of ZOD suppression using (a) the proposed method employing (b) a nonlocal metasurface based on guided mode resonance. (c) The corresponding 2D angle‑dependent transmittance. (d) Conventional ZOD suppression using a 4\textit{f} setup.}
\end{figure}

Figure 1(a) presents conceptual diagrams of the proposed ZOD‑free method using a nonlocal metasurface as a ZOD-stop.
Referring to Fig. 1, we provide a mathematical description of the ZOD and present the principle and concept of the proposed method.
Although the following explanation assumes the use of a phase-only SLM for generating target optical wavefronts, the same concept is applicable to DOEs and metasurfaces as well.

The phase-only SLM displays a phase pattern $\phi_s(x,y)$ to generate a target optical wavefront and is illuminated by a plane wave.
The SLM exhibits imperfections originating from pixel‑to‑pixel gaps, surface reflections, non‑uniformity, and fringing‑field effects, which give rise to the ZOD \cite{Palima2007,Moser2019,Engstrom2013}.
Consequently, the reflected light from the SLM contains both the signal component modulated by the phase pattern $\phi_s(x,y)$ and the noise component corresponding to the ZOD.
Let $a_n\exp(i\phi_n)$ denote the noise component, and the complex amplitude distribution of the optical wavefront immediately after the SLM can then be expressed as
\begin{equation}
u(x,y) =
 A(x,y) \{ \exp[i\phi_s(x,y)] + a_n(x,y) \exp[i\phi_n(x,y)] \},
\end{equation}
where $A(x,y)$ is an aperture function to support the effective area of the optical field.
Because the modulation area of an SLM typically has a rectangular shape, the illumination region must be restricted accordingly to minimize the generation of unmodulated light.
Therefore, by using a $L_x \times L_y$ rectangle function for the aperture function $A(x,y)$, Eq. (1) can be rewritten as follows:
\begin{equation}
u(x,y) =
 \textrm{rect} \left( \frac{x}{L_x},\frac{y}{L_y} \right) \{ \exp[i\phi_s(x,y)] + a_n(x,y) \exp[i\phi_n(x,y)] \}.
\end{equation}
The amplitude $a_n(x,y)$ of the noise component is much smaller than that of the signal component.
Moreover, the spatial variation of the phase $\phi_n(x,y)$ is sufficiently small.
Thus, the ZOD behaves as a plane‑wave‑like component and manifests as a faint background.

In the proposed method, the ZOD is suppressed by placing a nonlocal metasurface immediately after the SLM, as shown in Fig. 1(a).
Figure 1(b) illustrates a schematic cross‑section of the metasurface along the $x$‑direction, which consists of a high‑index, subwavelength‑period line‑and‑space structure formed on a low‑index substrate.
Such subwavelength patterning can be achieved using interference lithography \cite{Sahoo2023} or Talbot‑effect lithography \cite{Solak2011}, and, as described later, can also be produced by nanoimprint lithography.
By appropriately setting the structural parameters of the metasurface, namely the grating period $p$, duty cycle $w$, ridge height $d$, and residual‑layer thickness $t$, GMR can be excited.
When the GMR condition is satisfied, an incident plane wave impinging on the surface grating is partially diffracted, and the diffracted component couples into a guided wave supported by the high‑index residual layer.
Destructive interference between the leakage radiation of this guided wave and the undiffracted leakage component from the grating reduces the transmitted intensity.
As described in Subsection 3.1, proper tuning of the structural parameters allows the metasurface to exhibit the 2D angle‑dependent transmittance profile shown in Fig. 1(c).
This response indicates that normally incident light ($\theta_x \approx 0^\circ$), or ZOD, can be selectively attenuated.
Note that because the metasurface used in the proposed method is a 1D periodic structure, a distinctive angular selectivity appears only for incident angles $\theta_x$ varied in the direction parallel to the grating vector, as shown in Fig. 1(c).

To explain how the metasurface acts on the ZOD, and to contrast it with the conventional 4\textit{f}-based ZOD‑suppression approach (Fig. 1(d)), we consider the Fourier spectrum of the optical field $u(x,y)$ in Eq. (2).
\begin{equation}
U(\mu,\nu) =
 L_x L_y \textrm{sinc}(L_x\mu, L_y\nu) \otimes [ S(\mu,\nu) + N(\mu,\nu) ],
\end{equation}
where
\begin{equation}
S(\mu,\nu) =
 \iint_{-\infty}^{\infty} \exp[i\phi_s(x,y)] \exp[-i2\pi(x\mu+y\nu)] \, dx\, dy,
\end{equation}
\begin{equation}
N(\mu,\nu) =
 \iint_{-\infty}^{\infty} a_n(x,y) \exp[i\phi_n(x,y)] \exp[-i2\pi(x\mu+y\nu)] \, dx\, dy.
\end{equation}
$\otimes$ denotes the convolution operator.
Since the ZOD behaves as a plane‑wave‑like component, its Fourier spectrum $N(\mu,\nu)$ is well approximated by a delta function.
According to Eq. (3), the optical field component excluding the signal term is obtained as the convolution of the sinc function and the delta-like spectrum $N(\mu,\nu)$.
Thus, the ZOD spectrum is formed as a very strong peak with cross-shaped side lobes.
When scalar diffraction theory and the paraxial approximation hold, the 2D angular dependence $(\theta_x, \theta_y)$ becomes equivalent to the spatial‑frequency dependence $(\mu, \nu)$.
Therefore, the proposed method using the metasurface shown in Fig. 1(a) achieves functionality comparable to the spatial‑frequency filtering performed by the conventional 4\textit{f} setup shown in Fig. 1(d).
In the conventional method, the first lens performs a Fourier transform to obtain the spatial‑frequency spectrum, and by placing a mask at this Fourier plane, the concentrated delta‑function‑like zeroth‑order light can be removed.
Subsequently, the second lens performs another Fourier transform to generate the target wavefront without the zeroth‑order component.
However, the conventional approach requires two lenses, resulting in a bulky optical system.
In contrast, the proposed method employs a much simpler configuration.
Because no lenses are used, the proposed method benefits from the absence of lens‑induced aberrations and has no limitations imposed by the spatial‑frequency bandwidth of the relay lenses.


\section{Design, fabrication, and characterization of nanoimprinted nonlocal metasurface}
We designed and fabricated the nonlocal metasurface for the proposed method presented in Section 2 by employing a nanoimprint lithography.
Nanoimprint lithography offers excellent scalability and high throughput, making it a promising technique for realizing practical large‑area metasurfaces \cite{Oh2026,Yoon2020,Miyata2022}.
In this section, we describe the detailed design and fabrication procedure of the nanoimprinted nonlocal metasurface.
We also present the measured wavelength‑dependent and angle‑dependent transmittance characteristics of the fabricated samples.

\begin{figure}[b]
\centering\includegraphics[width=12cm]{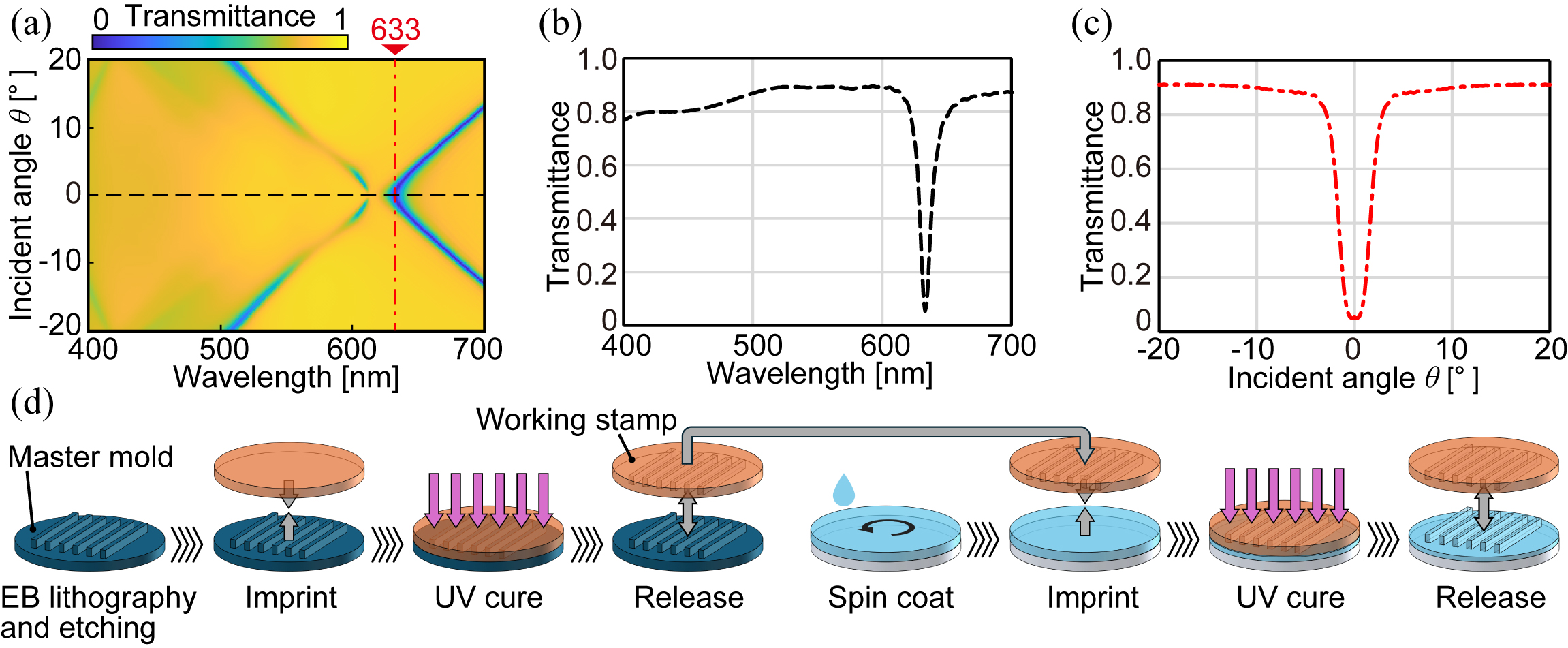}
\caption{Design and fabrication of the nanoimprinted nonlocal metasurface. (a) Angle‑ and wavelength‑dependent transmittance of the designed metasurface. (b) Wavelength- and (c) angle‑dependent transmittance profiles along the dashed and dash‑dotted lines in (a).
(d) Fabrication flow based on nanoimprint lithography.}
\end{figure}

\subsection{Design and fabrication}
Assuming the use of nanoimprint lithography, we designed a resin‑based nonlocal metasurface.
As the nanoimprint material, we employed a resin containing TiO\textsubscript{2} nanoparticles, which are essential for increasing the refractive index and thereby enabling effective GMR behavior.
We measured its complex refractive index using an ellipsometer.
The typical complex refractive indices were 1.918+0.009$i$ and 1.896+0.007$i$ at wavelength of 550 and 633 nm, respectively.
Based on the measured optical constants, we investigated device geometries that exhibit GMR at a wavelength of 633 nm using finite‑difference time‑domain (FDTD) simulations of Ansys Lumerical FDTD.
During the simulations, we assumed TM‑polarized illumination, meaning that the grating vector of the metasurface and the electric‑field vector lie in the same plane.
As a result of these simulations, we determined that the structure shown in Fig. 1(b), with a grating period of $p=$ 400 nm, duty cycle of $w=$ 225 nm, ridge height of $d=$ 190 nm, and residual‑layer thickness of $t=$ 130 nm.
Figure 2(a) presents the wavelength‑ and angle‑selective transmittance of the designed structure calculated using FDTD simulations. Two linearly dispersive dips are found in the transmittance map, which cross around $\lambda$ = 630 nm at normal incidence. These two bands are waveguide mode with $\pm$1 diffraction orders.
The cross‑sectional transmittance profiles along the dashed and dash‑dotted lines in Fig. 2(a) are presented in Figs. 2(b) and 2(d), respectively.
These results confirm that the wavelength at which the transmittance reaches its minimum under normal incidence ($0^\circ$) is 633.2 nm, and the angular full width at half maximum (FWHM) of the selectivity at the wavelength of 633 nm is 3.36$^\circ$.

The metasurface was fabricated by nanoimprint lithography (EVG6200TBN EVG, Austria). The nanoimprint Si master mold was separately prepared by the combination of EB lithography (F7000S-KYT-01, Advantest, Japan) and ICP etching (RIE-800iPB-KU, Samco, Japan). The surface morphology of the master mold was duplicated as the working stamp for the imprint process. A high-refractive-index resist (HN743, NTT-AT) was first spincoated on the glass substrate (Tempax, Shott, Germany) and baked at 110$^\circ$C for 30 sec. Then the resist was nanopatterned by nanoimprint with the working stamp under UV irradiation.

\begin{figure}[b]
\centering\includegraphics[width=12cm]{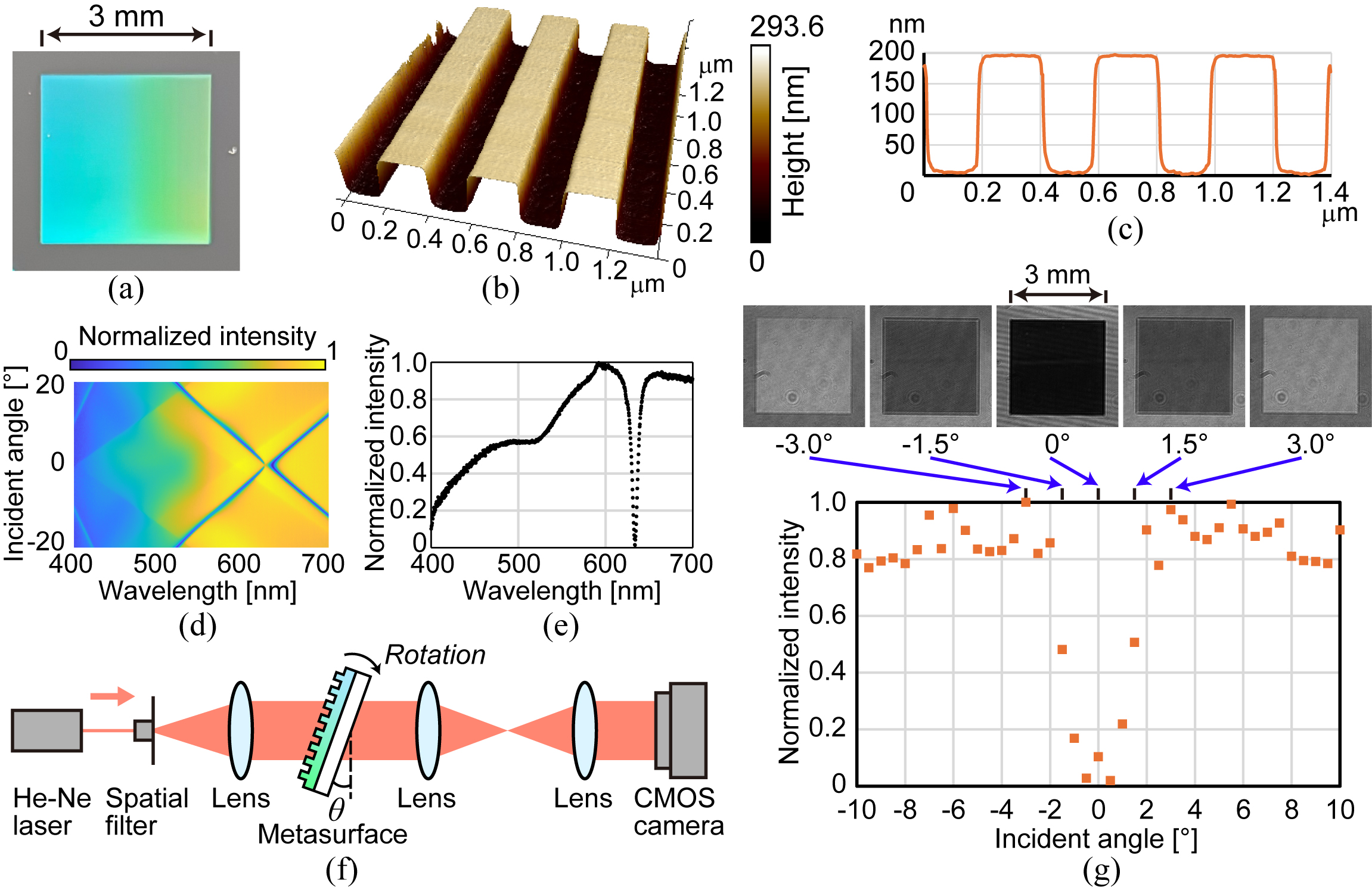}
\caption{Characterization of the fabricated metasurface.
(a) Photograph of the fabricated sample.
(b) 3D structure measured using an atomic force microscope together with the corresponding (c) cross‑sectional profile at the center.
(d) Measured wavelength‑ and angle‑dependent transmittance.
(e) Wavelength‑dependent transmittance for normal incidence ($0^\circ$).
(f) Optical setup for evaluating the in‑plane angular selectivity of the sample, and
(g) the corresponding evaluation results.}
\end{figure}

\subsection{Characterization}
Figure 3(a) shows a photograph of the fabricated 3‑mm‑square metasurface.
Figures 3(b) and 3(c) present the grating structure on the sample measured using an atomic force microscope (AFM). These results confirm that the intended grating structure was successfully fabricated.
The dispersion of the optical resonances was measured by a home-made variable-angle transmittance set-up. The light from a halogen lamp was polarized through a linear polarizer and incident on the sample placed on the rotation stage, and the zeroth-order transmission was measured by a spectrometer (SR4, Ocean Insight, USA) as a function of angle-of-incidence. 
Figure 3(d) shows the measured wavelength‑ and angle‑dependent transmittance.
Figure 3(e) presents the wavelength dependence for normally incident light ($\theta_y = 0^\circ$).
The minimum transmittance occurs at 633.96 nm, which agrees well with the FDTD simulation result shown in Fig. 2(b).
To evaluate the angular selectivity and to examine whether the response depends on the lateral position on the device, we measured the transmittance distribution immediately after the metasurface using the imaging setup shown in Fig. 3(f).
A plane wave from a 633 nm He–Ne laser was incident on the metasurface, which was mounted on a rotation stage and rotated in 0.5$^\circ$ increments.
The transmitted intensity images were recorded with a CMOS camera with 2048 $\times$ 2048 pixels and a pixel pitch of 6.5 $\mu$m.
Typical captured images and the resulting angle‑dependent transmittance are shown in Fig. 3(g).
For plotting the transmittance, the intensity values were averaged only over the region containing the grating structure in the captured images and normalized.
The FWHM of the angular response is 3.5$^\circ$, showing good agreement with the simulation result in Fig. 2(c).
These characterization results confirm that the intended metasurface was successfully fabricated.

\section{Proof-of-principle experiments}
In this section, we experimentally verify the principle of ZOD suppression using the 3‑mm‑square metasurface fabricated and characterized in Section 3.
We conducted experimental validation for two representative application scenarios.
The first corresponds to the use of the metasurface in conjunction with a surface‑relief DOE.
For a DOE in which the desired wavefront $S(\mu, \nu)$ and the ZOD $N(\mu, \nu)$ are completely separated in the spatial‑frequency domain, we apply the metasurface and evaluate its ZOD‑suppression performance in the Fraunhofer region.
The second scenario concerns holographic projection using an SLM.
In this case, the desired wavefront $S(\mu, \nu)$ and the ZOD $N(\mu, \nu)$ overlap in the spatial‑frequency domain, such that applying the metasurface inevitably attenuates a portion of the signal component as well. 
Nevertheless, we demonstrate that the metasurface can still effectively suppress the ZOD during wavefront reconstruction in the Fresnel region.
Finally, we present results obtained using a prototype 20‑mm‑square large‑area metasurface applied to 3D holographic projection.

\subsection{Surface-relief diffractive optical element}
We applied the fabricated metasurface to a surface‑relief DOE, as shown in Fig. 4.
Figure 4(a) shows the phase distribution of the target surface‑relief DOE, which consists of a binary phase value, 0 or $\pi$, forming a checkerboard arrangement.
The minimum feature size of this phase distribution is $5 \times 5~\mu\mathrm{m^2}$.
Such a checkerboard‑type phase grating is used in fields such as phase‑contrast imaging \cite{Morimoto2015,Liebel2020}, interferometer \cite{Yang2016,Nobukawa2022} and super-resolution imaging \cite{Bon2018}.
When Eq. (4) is evaluated on the optical axis $(\mu, \nu) = (0, 0)$, the 2D integral becomes nearly zero.
Therefore, ideally, as shown in the simulation result in Fig. 4(e), the Fourier spectrum of the signal component does not contain a zeroth-order component.

\begin{figure}[htbp]
\centering\includegraphics[width=6.5cm]{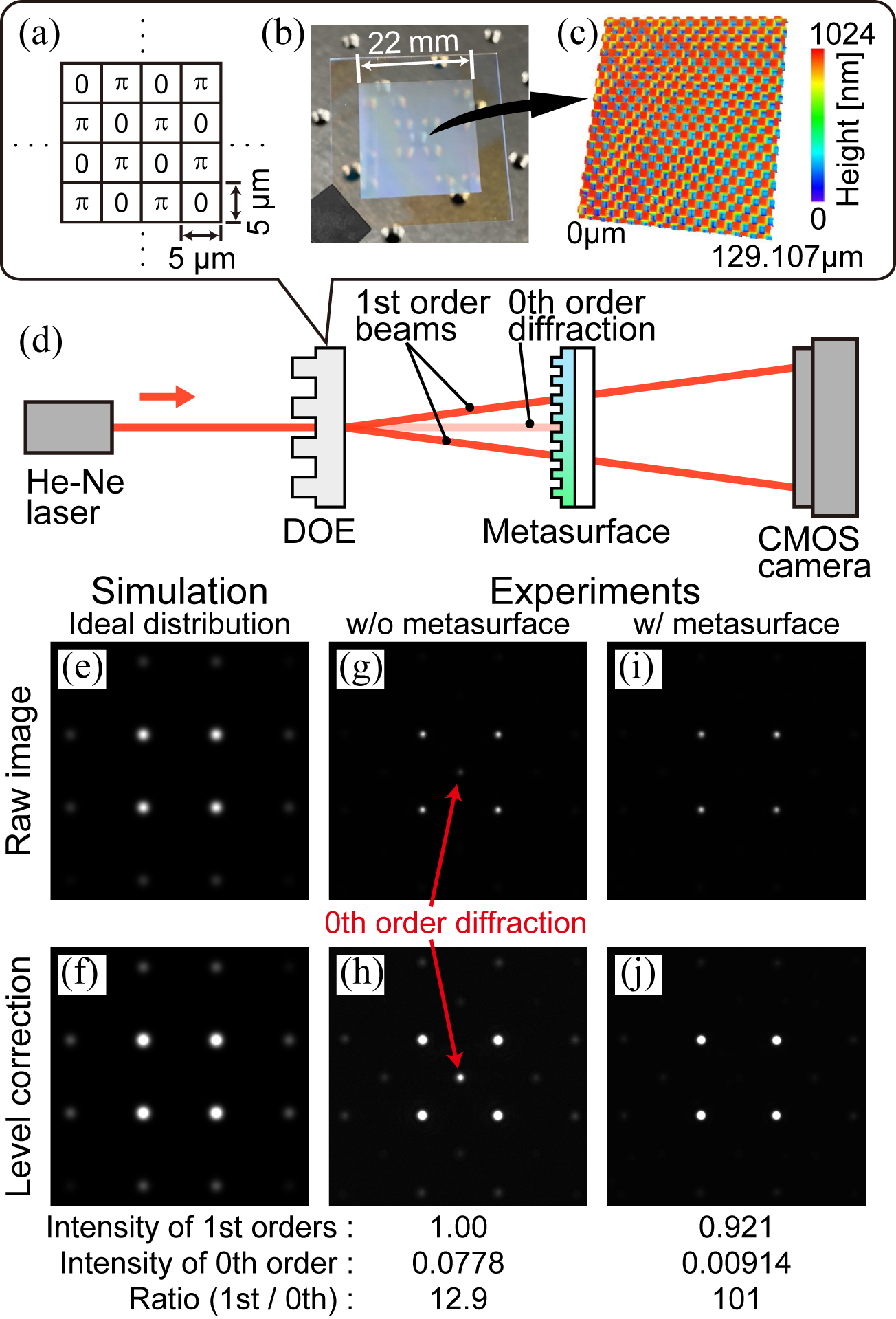}
\caption{
Experimental validation using a surface‑relief DOE.
(a) Ideal phase distribution of the designed surface‑relief DOE.
(b) Photograph of the fabricated element.
(c) 3D surface profile measured with a confocal laser microscope.
(d) Experimental setup.
(e, f) Simulated diffraction pattern produced by the ideal element.
Experimentally obtained diffraction patterns (g, h) without and (i, j) with the metasurface.
}
\end{figure}

The appearance of the surface‑relief DOE fabricated using photolithography and etching, together with the result of 3D measurement obtained by a confocal laser microscope, is shown in Figs. 4(b) and 4(c), respectively.
Fused silica was used as the substrate material of the surface‑relief DOE, and the height of the step required to generate a $\pi$ phase difference is 692 nm.
In contrast, the step height at the center region of the DOE shown in Fig. 4(c) is 698 nm, corresponding to a fabrication error of 0.8$\%$.
This fabrication error results in the appearance of the ZOD at the origin of the Fourier spectrum, as described later.
We constructed the optical setup shown in Fig. 4(d) and illuminated the surface‑relief DOE with a Gaussian beam at a wavelength of 633 nm, whose diameter was sufficiently larger than the minimum pitch 5 $\mu$m of the phase distribution.
The intensity distribution of the diffracted light in the Fraunhofer region was captured using an image sensor with 14192 $\times$ 10640 pixels and a pixel pitch of 3.76 $\mu$m.

Figure 4(g) shows the intensity distribution of the diffracted wave obtained without applying the metasurface.
For reference, a level‑corrected version of this intensity image is shown in Fig. 4(h).
Compared with the ideal spectrum shown in Figs. 4(e) and 4(f), the presence of the ZOD can be observed.
Figures 4(i) and 4(j) show the intensity distribution of the diffracted wave and its level‑corrected image, respectively, acquired when the metasurface was placed immediately after the DOE in the experimental setup.
It is evident that the ZOD is suppressed by applying the metasurface.
To quantitatively evaluate the reduction of the ZOD, we calculated the ratio of the total first‑order diffraction intensity, corresponding to the four main fundamental spatial frequencies, to the zeroth‑order intensity using the raw intensity data.
The evaluation results are shown at the bottom of Fig. 4.
The first‑order diffraction intensity obtained without the metasurface was normalized to 1.00.
With the application of the metasurface, the intensity of the first‑order signal component decreases to 0.921, which is attributable to the attenuation of incident waves at angles other than 0$^\circ$, as also suggested in the evaluation of Fig. 3(g).
However, the intensity of the zeroth‑order light is reduced from 0.0778 to 0.00914 with the introduction of the metasurface.
Since the ratio of first‑order to zeroth‑order intensity is 12.9 without the metasurface and 101 with it, the effectiveness of the proposed metasurface has been confirmed.

\subsection{Holographic projection with spatial light modulator}
In the previous subsection, we verified the effectiveness of the proposed metasurface in a case where the signal spectrum and the ZOD are completely separated, and in the Fraunhofer region.
In this subsection, we conduct experiments in a holographic projection application, which represents a case where the signal spectrum and the ZOD overlap.
We applied the proposed metasurface to the Fresnel-region wavefront generated by an SLM and evaluated the effectiveness of the method.

\begin{figure}[t]
\centering\includegraphics[width=6.5cm]{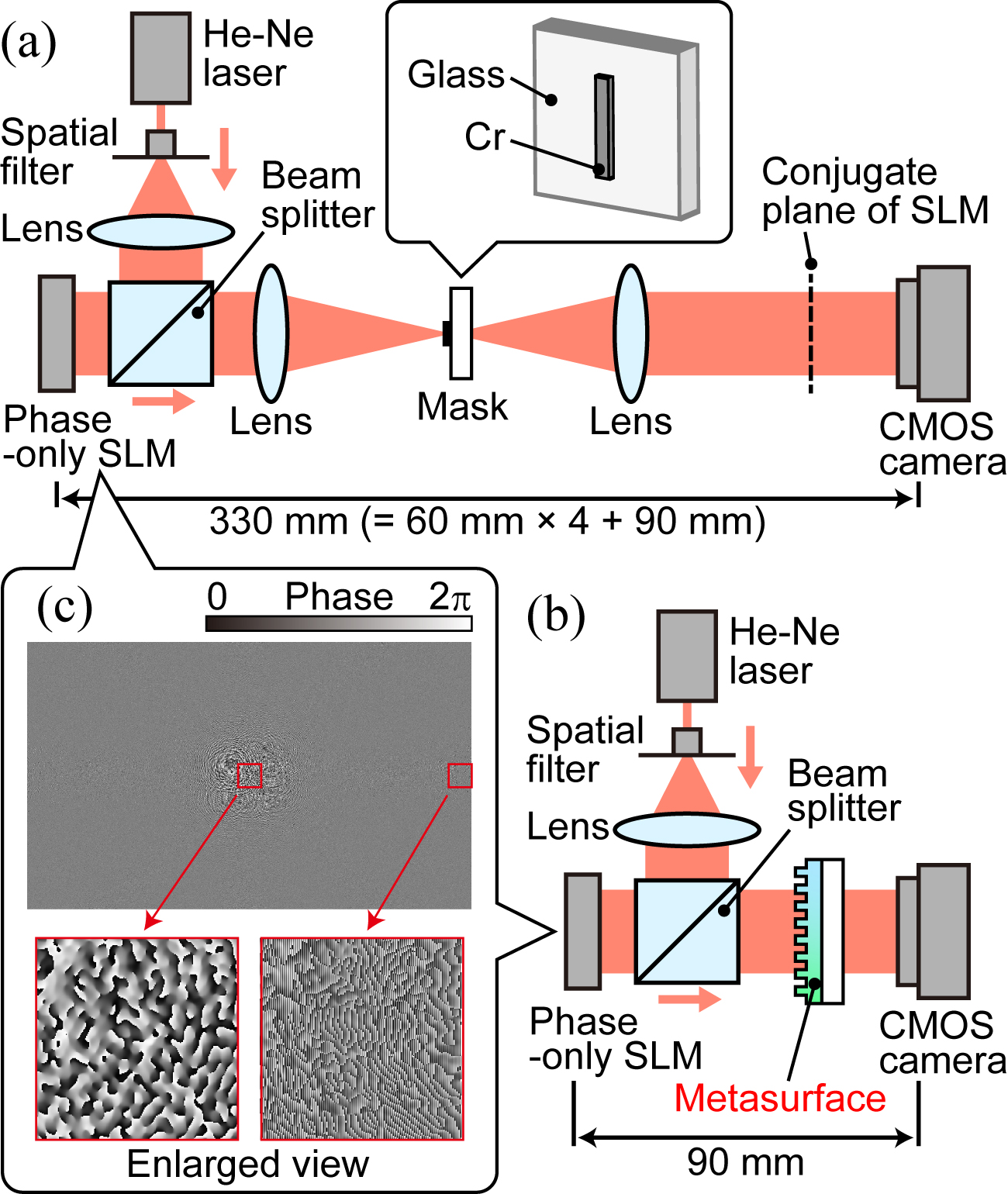}
\caption{Holographic projection system based on (a) 4\textit{f} and (b) free-space setups. (c) Example phase hologram displayed onto a phase-only SLM.}
\end{figure}

\begin{figure}[t]
\centering\includegraphics[width=13cm]{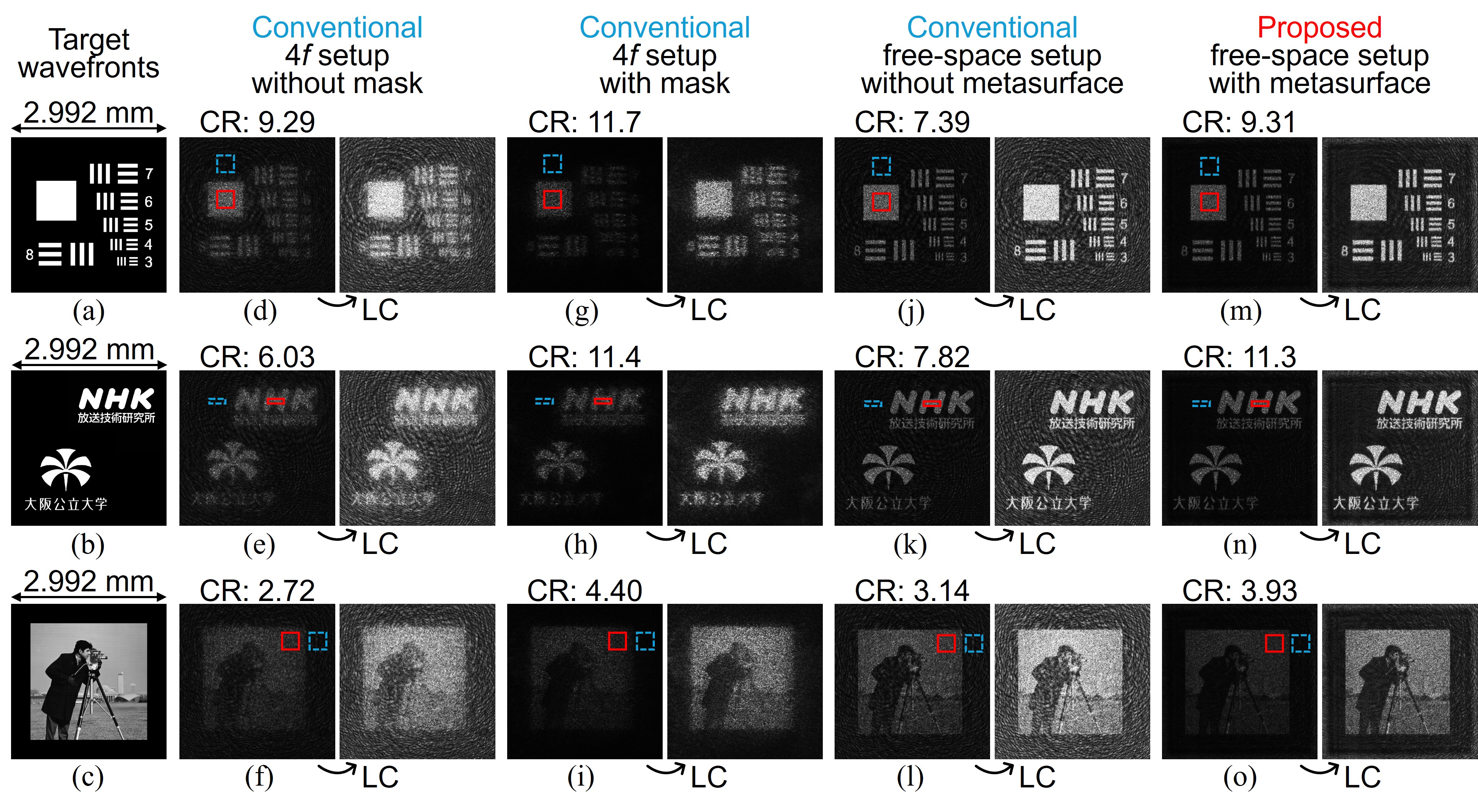}
\caption{Reconstruction results for four experimental conditions.
(a–c) Target wavefronts used for hologram design.
Reconstructed images obtained using the 4\textit{f} setup (d-f) without and (g-i) with a mask.
Reconstructed images obtained using the free‑space setup (j-l) without and (m-o) with the metasurface.
For clarity, level‑corrected (LC) versions of each reconstructed image are also shown.
The evaluated contrast‑ratio (CR) values are displayed at the top of each reconstructed image.}
\end{figure}

To compare the conventional method with the proposed method, two optical setups shown in Figs. 5(a) and 5(b) were used.
As the SLM, we employed a reflective, liquid-crystal-on-silicon phase modulator with 4160 $\times$ 2464 pixels and a pixel pitch of 3.74 $\mu$m.
The target wavefronts to be generated were shown in Figs. 6(a-c).
These image sizes were set to fit within the 3 $\times$ 3 mm area of the fabricated metasurface.
Figure 5(c) shows an example phase hologram to generate the target shown in Fig. 6(a).
From each target, a phase hologram was designed by an iterative method in a Fresnel domain based on the Gerchberg--Saxton algorithm \cite{Blinder2022}, or Fresnel ping--pong algorithm \cite{Dorsch1994}.
In the setup for the conventional method, a 4\textit{f} setup was constructed using two lenses with an effective diameter of 25.4 mm and a focal length of 60 mm.
To suppress the ZOD, we fabricated a Cr-coated glass mask and placed it at the Fourier plane of the 4\textit{f} setup.
The width of the light‑blocking region was set to 3.66 mm to approximately match the angular selectivity of the metasurface shown in Fig. 3(g).
In contrast, in the proposed setup, we placed the fabricated sample in front of a CMOS camera.
A CMOS camera with 2048 $\times$ 2048 pixels and a pixel pitch of 6.5 $\mu$m was placed 90 mm away from the conjugate plane of the SLM.
For comparison, we captured reconstructed images under four conditions: using the 4\textit{f} setup in Fig. 5(a) both without and with a mask, and using the free‑space setup in Fig. 5(b) both without and with the metasurface.

Figure 6 shows the results.
The reconstructed images were captured with exposure adjusted to avoid saturation.
For ease of interpretation, the level‑corrected versions of the reconstructed images are additionally shown in Fig. 6.
To quantitatively compare the presence of ZOD in each reconstructed image, we evaluated the image quality using a contrast ratio (CR), defined as the intensity ratio between the originally bright region (indicated by the solid‑line rectangle) and the originally dark region (indicated by the dashed‑line rectangle).
In Fresnel hologram reconstruction experiments, the ZOD appears as background noise, and thus the CR serves as an effective metric for evaluating the ZOD influence.
Figures 6(d-f) show that When using the 4\textit{f} setup, the reconstructed images inevitably suffer from distortion due to lens aberrations and the inherent bandwidth limitation.
Although applying the mask improves the CR for all reconstructed images and effectively suppresses the ZOD, the effects of aberrations and bandwidth limitation remain, as shown in Figs. 6(g–i).
As demonstrated in Figs. 6(j–l), the free‑space optical setup in Fig. 5(b) does not suffer from lens aberrations or spatial‑frequency bandwidth limitations.
However, without the metasurface, the ZOD appears and CRs are low.
Finally, in the results obtained with the metasurface, shown in Figs. 6(m–o), the ZOD is suppressed, and the CR is higher than that in Figs. 6(j–l).
Note that the CR achieved by the proposed method (Figs. 6(m–o)) is lower than that of the conventional mask-based method (Figs. 6(g–i)).
As indicated by the angular selectivity of the proposed metasurface shown in Fig. 3(g), this is because its transmission is not reduced to zero.
Further reducing the transmission would require optimizing the device structure or stacking multiple metasurface layers.
Conversely, introducing an off‑axis carrier into the phase hologram is also effective \cite{Zhang2009}. Although this reduces the available spatial‑frequency bandwidth, it prevents attenuation of the signal by the metasurface and is therefore expected to improve the CR.
These directions fall outside the scope of the present work and remain subjects for future investigation.
From these experimental results, we demonstrated that the proposed method allows ZOD suppression using a simple optical setup without requiring a 4\textit{f} setup, thus eliminating the influence of lens aberrations and bandwidth limitations.

Finally, to demonstrate the scalability of nanoimprint lithography, we fabricated a 20‑mm‑square metasurface on a 4‑inch glass substrate using the same process conditions as those in Fig. 2(d).
Its appearance is shown in Fig. 7(a).
The fabricated metasurface is sufficiently larger than the SLM area of $9.215 \times 15.558~\mathrm{mm}^2$ shown in Fig. 7(b).
We applied this metasurface in the setup shown in Fig. 5(b) and conducted verification experiments that generate large reconstructed images from phase holograms to evaluate ZOD suppression.
The phase hologram used in this experiment was designed using the iterative method to reconstruct two images located at 90 mm and 110 mm, as shown in Fig. 7(b).
The reconstructed images without and with applying the metasurface are shown in Figs. 7(c) and 7(d), which were captured by displacing the CMOS camera along the optical axis.
All images were recorded with the same exposure time.
Note that level correction was not applied to the reconstructed images in Figs. 7(c) and 7(d) because their reconstructed image size is larger than that in Fig. 6, resulting in a lower optical intensity per unit area and fewer sharp intensity peaks.
The evaluated CR values are provided at the top of each reconstructed image.
With the metasurface applied, the CR of both the front and rear reconstructed images is improved.
As can be seen from the upper and lower regions of the reconstructed images in Figs. 7(c) and 7(d), the proposed metasurface also effectively suppresses reflected light originating from the non‑modulated region of the SLM.
These experiments demonstrate that large metasurfaces can be fabricated using nanoimprint lithography and that the proposed method can suppress ZOD even for large 3D reconstructed images.

\begin{figure}[t]
\centering\includegraphics[width=11cm]{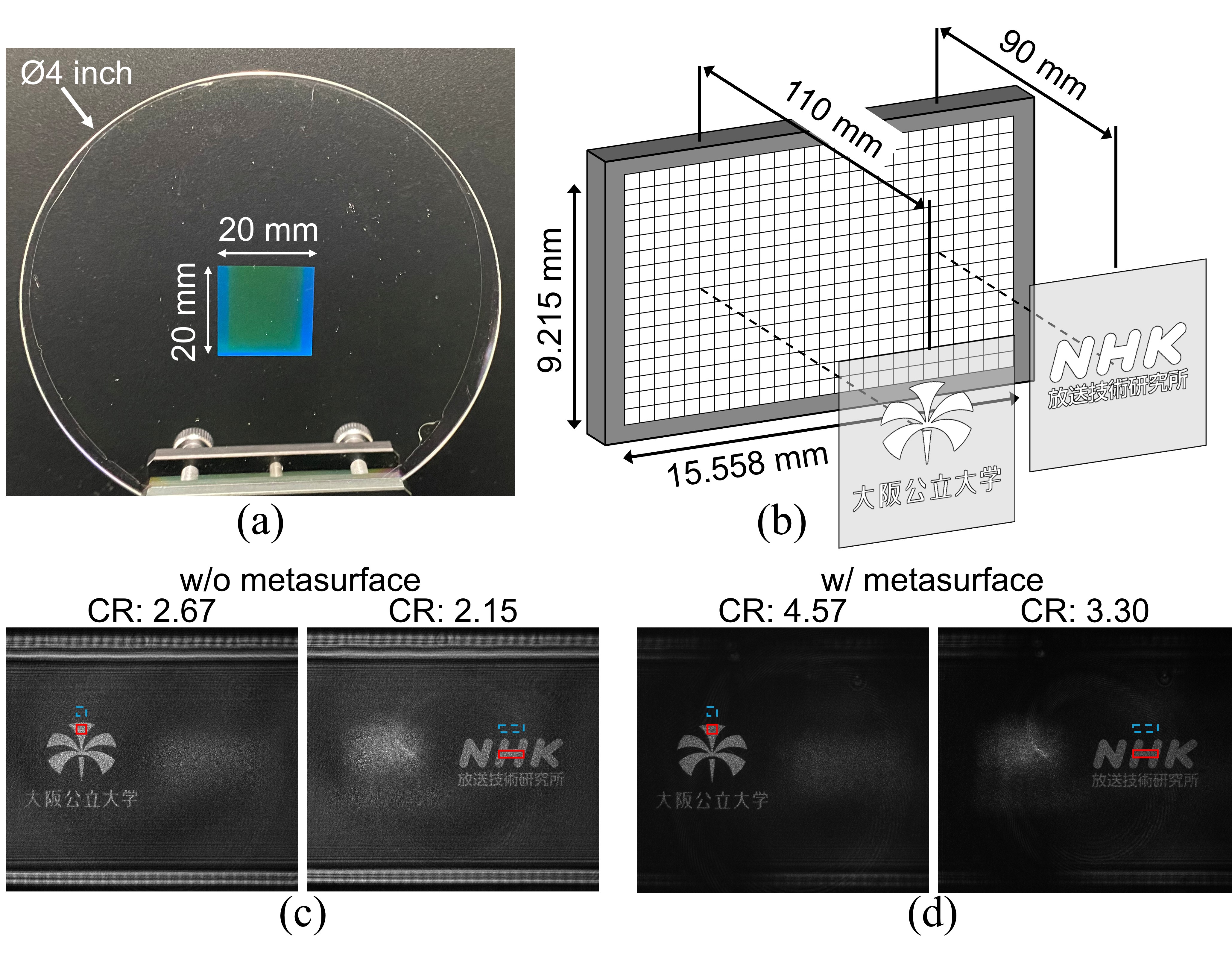}
\caption{
Experimental demonstration using a 20‑mm‑square metasurface.
(a) Photograph of the fabricated sample.
(b) Arrangement of the SLM and the target wavefronts.
Reconstructed images obtained (c) without and (d) with the metasurface applied.}
\end{figure}

\section{Conclusions}
We proposed a method for suppressing the zeroth‑order diffraction (ZOD) using a nanoimprinted nonlocal metasurface.
The effectiveness of the proposed approach was experimentally demonstrated in two representative applications: a surface‑relief DOE and a holographic projection system based on an SLM.
The results show that the proposed method enables high‑quality wavefront generation with a compact optical configuration that is free from lens‑induced aberrations and spatial‑frequency bandwidth limitations.

The metasurface employed in our method consists of a simple 1D line‑and‑space structure, which is relatively easy to fabricate.
Although nanoimprint lithography typically leaves a residual layer that can be problematic for metasurfaces \cite{Dirdal2020, Park2025}, the proposed metasurface intentionally requires a uniform layer beneath the nanostructures to excite GMR.
In this sense, the residual layer produced by nanoimprint lithography naturally fulfills this requirement, making the fabrication process highly compatible with the metasurface design.
Furthermore, we validated this capability by prototyping a 20‑mm‑square metasurface on a 4‑inch glass substrate and demonstrating its performance in a practical experimental setup.
These results collectively demonstrate the scalability of the proposed method and its potential for large‑area fabrication and mass production.

In this study, the metasurface material was based on a resin mixed with TiO\textsubscript{2} nanoparticles.
While TiO\textsubscript{2} can induce photocatalytic degradation of the resin, previous studies \cite{Ha2025} have shown that encapsulation techniques that exclude oxygen and moisture can effectively mitigate this issue, suggesting that the material challenge is technically solvable. 
Although we focused on a 1D line‑and‑space structure for manufacturability and ease of use, extending the design to 2D periodic structures would enable directional selectivity in full 2D angular space, and more sophisticated metasurface designs could further provide additional functionalities \cite{Pearson2025}.

Overall, the proposed metasurface‑based ZOD‑suppression technique offers a compact and lightweight solution to a longstanding challenge in diffractive optics.
Because it can be applied to a wide range of optical systems employing DOEs or SLMs, we expect the approach to have broad impact across optical engineering.
In particular, it holds strong potential for advancing holographic near‑eye displays in augmented‑reality and virtual‑reality applications \cite{Chang2020,He2019}.


\begin{backmatter}

\bmsection{Acknowledgment}
We thank Dr. TienYang Lo at Kyoto University for the experimental support. 

\bmsection{Data Availability Statement}
Data underlying the results presented in this paper are not publicly available at this time but may be obtained from the authors upon reasonable request.


\end{backmatter}


\end{document}